\let\TeXyear\year
\documentclass{pramana}
\let\setyear\year
\let\year\TeXyear

\usepackage{graphicx,amsmath,bm}
\usepackage{hyperref}
\usepackage{indentfirst}
\usepackage{float}
\usepackage{amssymb}
\usepackage{mathtools} 
\usepackage{commath}
\usepackage{hyperref}
\usepackage{tikz,xcolor} 
\usepackage[noadjust]{cite}

\foreach \x in {A, ..., Z}{\expandafter\xdef\csname orcid\x\endcsname{\noexpand\href{https://orcid.org/\csname orcidauthor\x\endcsname}{\noexpand\orcidicon}}
}

\begin{document}

\title{Two-stream Plasma Instability as a Potential Mechanism for Particle\\ Escape from the Venusian Ionosphere}


\author{S. Dey\textsuperscript{1}, S. Ghosh\textsuperscript{2},
	D. Maity\textsuperscript{1}, A. De\textsuperscript{3}\and S. Chandra\textsuperscript{4,5*}}
\affilOne{\textsuperscript{1} Department of Physics, Visva-Bharati University, Santiniketan, Bolpur 731235, India\\}
\affilTwo{\textsuperscript{2} Department of Physics, Ramakrishna Mission Residential College, Narendrapur, Kolkata 700103, India\\}
\affilThree{\textsuperscript{3} Department of Physics, Indian Institute of Technology Madras, Chennai, 600036, India\\}
\affilFour{\textsuperscript{4} Department of Physics, Government General Degree College at Kushmandi, Dakshin Dinajpur 733121, India\\}
\affilFive{\textsuperscript{5}Institute of Natural Sciences and Applied Technology, Kolkata 700032, India}


\twocolumn[{

\maketitle

\corres{swarniv147@gmail.com}

\msinfo{19 August 2021}{2 June 2022}{5 July 2022}

\begin{abstract}
In this work we investigate the possibility of two-stream instability in the Venusian atmosphere to lead to momentum transfer to subsequent escape of Hydrogen and Oxygen ions from the ionosphere. We employ the hydrodynamic model and obtain the linear dispersion relation from which the two-stream instability is studied. Further the interaction of solar wind with the ions of Venus ionosphere from which the instability sets in, has been studied with the data from ASPERA-4 of Venus Express (VEX). The data supports the fact that the two-stream instability can provide sufficient energy to accelerate ions to escape velocity of the planet and thus leave the Venusian ionosphere.
\end{abstract}

\keywords{Two Stream Instability, Linear Dispersion, Solar Wind (SW), Hydrodynamic Model, Venus 
	Express (VEX), ASPERA-4}

\pacs{96.30.Ea; 94.20.wh; 52.35.Qz; 11.55.Fv; 52.25.Jm}

}]


\doinum{00.XXXX/XXXXXX-XXX-XXX-X}
\artcitid{\#\#\#\#}
\volnum{000}
\setyear{2022}
\pgrange{01-08}
\setcounter{page}{01}
\lp{8}

\section{Introduction}
\indent Planets that have strong magnetic fields in their interior region, like Earth, Jupiter, Saturn, and Mercury, are enclosed by invisible magnetosphere \cite{Ham}. The charged particles of the solar wind radiation (electrons and protons) are deflected by their magnetic fields as they stream far away from the Sun. This deflection by the magnetic field creates a magnetic sphere that acts as a protective "bubble" covering the planet \cite{zhang2012magnetic}. Venus has no intrinsic magnetic field to act as a protection against the incoming stream of charged particles. However, the solar wind and UV radiation from sun, pulls out electrons from the atoms and molecules within the higher atmosphere, making a section of electrically charged gas referred to as the ionosphere. The planet is protected partially by the induced magnetic force field due to this ionosphere. But this induced magnetic field is very small in magnitude and not sufficient to deflect the solar wind unlike other planets. Thus, the electrons in the solar wind end up directly interacting with the higher atmosphere. We have suggested a likely ion escape route resulting from the instability under some physical conditions. We have used analytical and simulation tools and their outcomes, which depend on the occurrences of the ionization processes, as well as solar wind conditions. Newborn ions are at rest relative to the Venus frame. From previous studies \cite{futaana2017solar}, \cite{pope2009giant} we can observe that ion acceleration on Venus can occur through three mechanisms: (a) ion uptake (through acceleration in a convective electric field), (b) through induced instabilities in the ionosphere, and lastly (c) through polarization electrical fields at low altitude on the night side of the Venusian ionosphere. We are going to explore the second mechanism of ion acceleration in this paper.

This paper explores the possibility of two-stream instability to cause ions to escape from the atmosphere of Venus. It has been observed that the critical condition under which the instability can occur for the classical regime plays a major role in particle energization. The following is the outline of the paper. We discuss the fluid model of Venusian plasma in section \ref{sec2}, the dispersion relation has been obtained in section \ref{disper}, section \ref{dis} finally describes the possible ion escape mechanisms via two-stream instability by analyzing the Venus Express data.

\section{Fluid model of the Venusian ionosphere}\label{sec2}
\noindent In 1937 Jeffreys \cite{jeffreys1937density} first built the model of Venus from Bullens' model \cite{bullen1950venus} of the Earth. In 1983, Zharkov \cite{zharkov1983models}  provided a complete model of Venus. Venus has an extremely low magnetic field. So planetary magnetosphere is less important in this case. A magnetospheric plasma shell which exists around some planets have been investigated by many authors \cite{goswami2020electron, sarkar2021growth, thakur2021stationary, chandra2021formation, sarkar2021forced, kapoor2022magnetosonic, chandra2022multistability} but not with respect to Venus. We have considered a homogeneous plasma \cite{das2020amplitude, das2020nonlinear} comprised of electrons and ions. The basic governing equations for the fluid model \cite{dey2021chaotic, samanta2021nonlinear, ghosh2021chaotic, paul2016w, chandra2020analytical, goswami2021amplitude, roychowdhury2021stationary, majumdar2021study, ghosh2021linear, ghosh2021propagation, goswami2022nonlinear} are as follows \cite{ren2008dispersion, kourakis2007nonlinear, sarkar2020resonant, jehan2010perpendicular, singh2018second}:
\begin{equation}
	\frac{\partial n_{j}}{\partial t}+\mathbf{\nabla}\cdot \left(n_{j}\mathbf{v_{j}}\right)=0 
\end{equation}

\begin{equation}
\label{eqn:E2}
	\left(\frac{\partial \mathbf{v_{j}}}{\partial t}+\mathbf{v_{j} \cdot\nabla} \mathbf{v_{j}}\right)= -\frac{Q_{j}}{m_{j}}\left(\mathbf{\nabla} \phi- \mathbf{v_{j}}\times \mathbf{B}\right)-\frac{1}{m_{j}n_{j}}\mathbf{\nabla}P_{j}
\end{equation}

\begin{equation}
	\mathbf{\nabla}\times \mathbf{E}=-\frac{\partial \mathbf{B}}{\partial t}
\end{equation}

\noindent The terms are on the right side of equation (\ref{eqn:E2}) specified as Lorentz force, stress term \cite{das2020effects}. Let us assume the equation of state \cite{ren2008dispersion} is given by:

\begin{equation}
	P_{j0} = n_{j0}k_{B}T
\end{equation}
where $n_{j}$ signifies the number density, $v_{j}$ represents the fluid velocity, $p_{j}$ indicates the pressure, $m_{j}$ represents the mass, $\mathbf{E}$ and $\mathbf{B}$ represents the electric and magnetic field respectively, and $k_{B}$ is the Boltzmann constant. 

\section{Dispersion Relation}\label{disper}
As the solar wind flows freely into the Venusian atmosphere, it excites the neutral particles to form ions. This constitutes the plasma of the ionsphere. The resulting magnetic field has a very low value unlike the intrinsic magnetic fields of planets like, Earth, Jupiter, etc. The electrons in the solar winds streams into the stationary ions of the ionsphere. This relative motion between the ions and fast moving electrons causes the so called two-stream instability. Two-stream instability can be discussed based on the consideration of electron-ion streams having densities $n_e, n_i$ and temperatures $T_e$, $T_i$ with the velocities $v_e, v_i$. The electron and ion thermal velocities are $v_{T,e}, v_{T,i}$. We further investigate whether the aforesaid situation is relevant with Venusian ionosphere or not. Here $\omega_{pi}$ and $\omega_{pe}$ represents the plasma frequency with which ions and electrons oscillate, respectively. We get the dispersion as follows:
\begin{equation}\label{FIRST_EQ}
		1-\frac{\omega_{pe}^2/4}{\left(\omega - kv_{e}\right)^2-k^{2}v_{T,e}^{2}/2}
		-\frac{\omega_{pi}^2/4}{(\omega - kv_{i})^2-k^{2}v_{T,i}^{2}/2} = 0
\end{equation}
$v_{T,e}=\sqrt{2T_e/m_e}$ and $v_{T,i}=\sqrt{2T_i/m_i}$ are the thermal speed of electrons and ions respectively \cite{ren2008dispersion}. The equation is quite same to the results in \cite{ren2008dispersion},\cite{chandra2013electron} and Hsia \cite{hsia1979generalized} discarding the thermal terms. We can simplify by normalizing the time and relative velocity (or drift velocity) as $1/\omega_{pe}$ and $\omega_{pe}/k$ respectively. Relative velocity, $v\equiv v_{e}-v_{i}$ can be obtain by a simple analysis where, $v_{e}$ and $v_{i}$ are oppositely directed. For electron-positron pair plasma it can be shown that purely oscillating two stream instability can be possible for $m_{e}/m_{p}=1$ ,where the relative velocity needs to be periodic $sin(\omega_{0}t)$. From Qin et al.\cite{qin2014two} we can clearly see that the growth rate has a dependence on
$\omega_{p}/\omega_{0}$ and $kv/\omega_{pe}$ in a very complex way with $\omega_{p}=\sqrt{2}\omega_{pe}$. If the driving frequency is slightly higher or lower than the plasma frequency, the oscillations become unstable.

\subsection{Conditions for onset of instability: }
We simplify the equation (5) in the rest frame of ions:
\begin{equation}
    \mathcal{F}(\omega) = \frac{\omega_{pe}^2/4}{\left(\omega - kv_{e}\right)^2-k^{2}v_{T,e}^{2}/2}+\frac{\omega_{pi}^2/4}{\omega^{2}-k^{2}v_{T,i}^{2}/2}
\end{equation}
For extremum point we have,
\begin{equation}\begin{split}
    \xi = &\left[q+\left\{q^{2}+\left(r-p^{2}\right)^{3}\right\}^{1/2}\right]^{1/3} +\\& \left[q-\left\{q^{2}+\left(r-p^{2}\right)^{3}\right\}^{1/2}\right]^{1/3}+p
    \end{split}
\end{equation}
Where, $\alpha=\frac{\omega_{pi}^2}{\omega_{pe}^2}, \xi=\frac{\omega}{kv_{e}} , p=\frac{2\alpha^{2}}{3(1+\alpha^{2})},\\ q=\frac{p^{3}+\left[(-2\alpha^{2})\left\{\alpha^{2}-1-\left(1+\alpha^{2}\right)\frac{v_{0}^{2}}{2v_{e}^{2}}\right\}-3\left(1+\alpha^{2}\right)\frac{v_{0}^{2}}{2v_{e}^{2}}\right]}{6\left(1+\alpha^{2}\right)^{2}},\\ r=\frac{\alpha^{2}-1-\left(1+\alpha^{2}\right)\frac{v_{0}^{2}}{2v_{e}^{2}}}{3\left(1+\alpha^{2}\right)}$ and $v_{0}$ is the speed at thermal equilibrium. Now at that point, 
\begin{equation}\begin{split}
    \omega \equiv \omega_{s}= &kv_{e}\left[q+\left\{q^{2}+\left(r-p^{2}\right)^{3}\right\}^{1/2}\right]^{1/3} +\\& kv_{e}\left[q-\left\{q^{2}+\left(r-p^{2}\right)^{3}\right\}^{1/2}\right]^{1/3} + pkv_{e}
    \end{split}
\end{equation}
The condition for instability, $\mathcal{F}(\omega_{s})>1$
\\
\\
If the above obtained value for $\omega_s$ is substituted in the expression for $\mathcal{F}(\omega)$ and is equated to 1, then by little algebraic manipulation, $v_e$ will be determined. We define this electron velocity satisfying the instability condition to be the threshold velocity($v_{th}$).
\begin{equation}
    v_{th} \equiv v=\frac{\omega_{pe}}{k}\left[1+\left(\frac{m_{e}}{m_{p}}^{1/3}\right)\right]^{3/2}
\end{equation}
Any relative electron velocity greater than this threshold velocity will make the system unstable and the resultant instability might accelerate the ions to escape velocity. Detailed discussion on this is done in the subsequent sections. 
\section{Discussion}\label{dis}
The particles have a speed near to the phase speed with which plasma waves can supply the energy necessary for escape. A probable process for coupling of collisionless momentum between the plasmas which are interpenetrating with each other at the blocking layer might result in a stream of electrons causing the streaming instability.

\begin{figure*}[ht]
	\centering{\includegraphics[height=5cm,width=17.5cm]{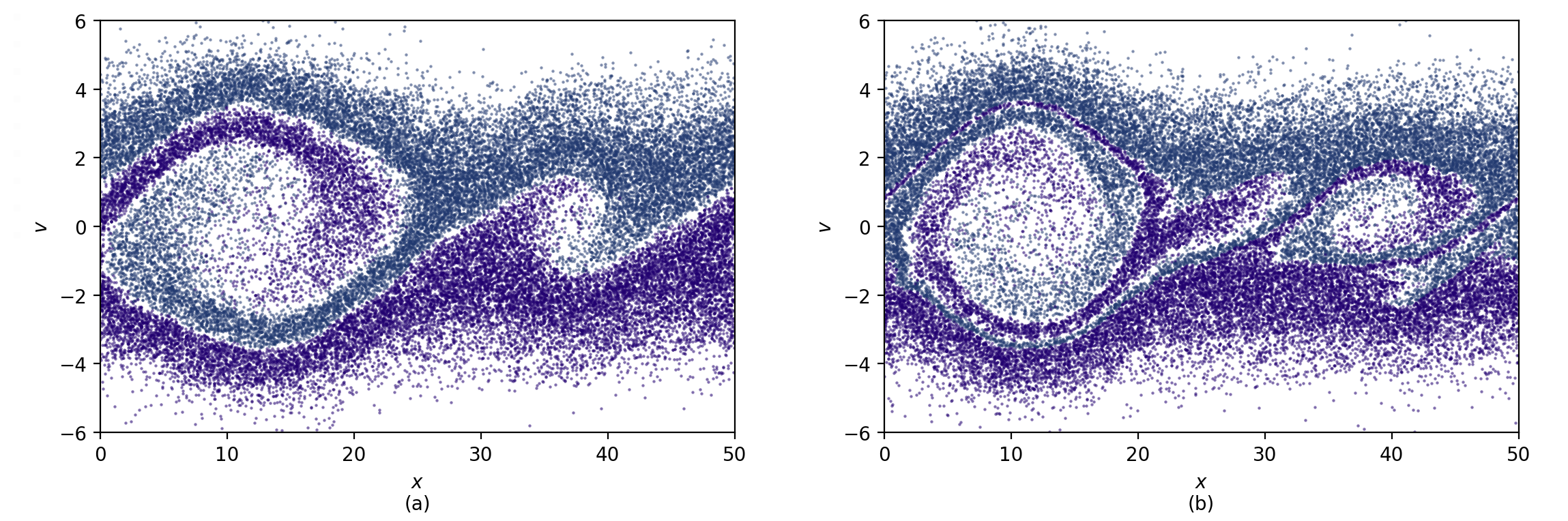}}
    \caption{Phase space particle-in-cell simulation of streaming instability occurring in (a) and (b) when the velocity of threshold ($v\equiv v_{th}$) is much lower than relative velocity. Captured for specific time values as indicated, (a) 20 s, (b) 30 s.}
    \label{Fig:2}
	
\end{figure*}

\begin{figure*}[ht]
	\centering{\includegraphics[height=5cm,width=17.5cm]{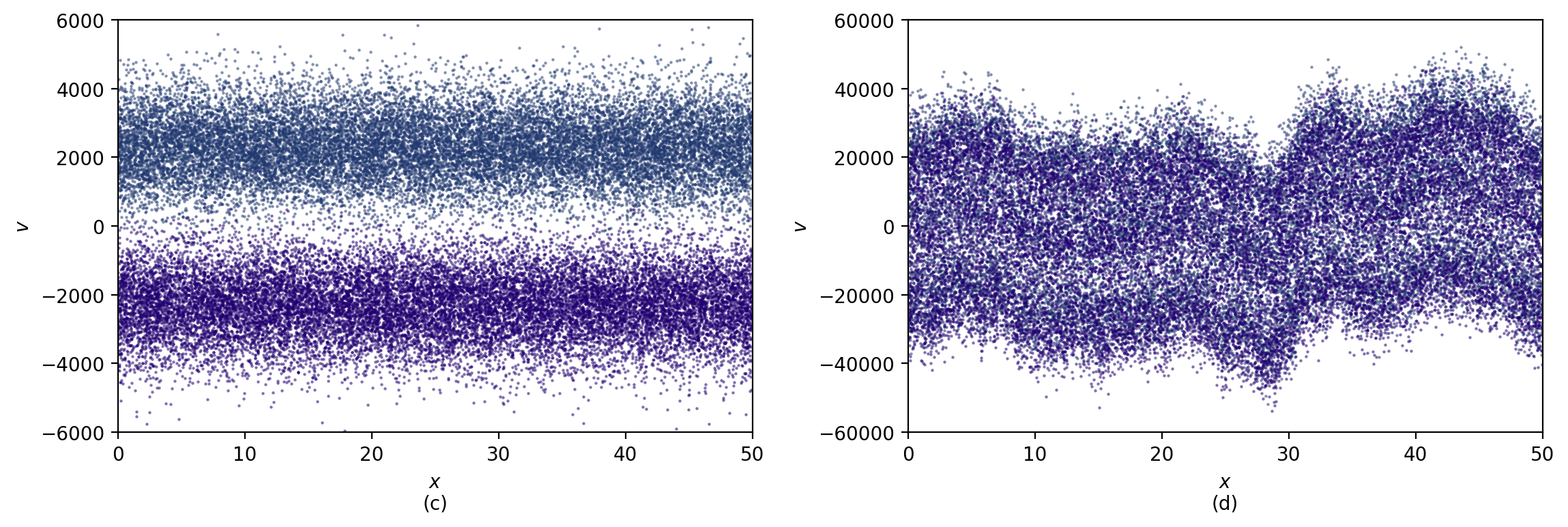}}
    \caption{Phase space particle-in-cell simulation of streaming instability occurring in (c) and (d) when the threshold velocity is large enough than the relative velocity. Captured for specific time values as indicated, (c) 20 s, (d) 30 s}
    \label{Fig:2a}
\end{figure*}

\subsection{Analysis with PIC simulation}
In this system, electrons got accelerated by the electric field in the vicinity of particle following the Poisson equation,
\begin{equation*}
    \frac{d^{2}\phi(x_{j})}{dx^{2}} \approx \frac{\phi_{j+1}+\phi_{j-1}-2\phi_{j}}{(\Delta x)^{2}}
\end{equation*}
Using the above equation, we looked for the potential at each point in the mesh. We assumed a periodic sinusoidal boundary condition and interpolated from adjacent grid-points $j$ and $j+1$. Using the leap-frog integration method, we calculated the velocity and position at each time step.

\subsubsection{Simulation Parameters: }
The parameter values for our configurations are as follows: the PIC algorithm were simulated with 40000 number of particles $(N)$ and 400 number of mesh cells. The growing of instability were measured with a time step $(dt)$ of 1 sec from $t=1$ to $50$. To construct the matrix for the gradient operator for computing the first derivative we used the minimum separation $\Delta x = 0.125$. To calculate the initial gravitational accelerations of electrons we have used those values. Next we constructed two oppositely moving streams adding the periodic perturbation with amplitude of sinusoidal function around 0.1. Initial relative velocity is taken as 4.5 km/s.
\subsubsection{Initial Conditions:}
In our code, we implemented two uniform counter-propagating streams travel initially (t=0). The orientations of the particles are chosen randomly with a uniform distribution. The velocities are measured using a Gaussian centered distribution about the rate of streams. A tiny perturbation in periodic form (sinusoidal function) is applied to the velocity distribution. We simulated with a large number of particles (around 40000 particles) in phase space.

\subsubsection{Emergent Phase-space:}

Let now discuss how the ions get energize by interpenetrating flow of particles. A prevalent electric field, which is not stable, links the two streams. The energy from the strongly interacting streams is transferred to the driven electric field. The streams mix as the destabilizing electric field increases and continue to act, ultimately destroying the streams or dispersing the particles in phase space. As the acceleration of the streams grows, the instability tends to increase, potentially causing the streams to become more turbulent. Several electrons have been ascertained moving at a faster rate than their initial velocity due to a boost in energy from the wave, continuing to increase their velocity at the outlay of the wave energy, leading to a reduction in wave velocity. The energy due to the electric field increases as the stream velocity increases. The energy of the particles increased as the density of the streaming particles increased, however, at different rates. Indicating that the streaming particles become thermalized across the simulation, causing the plasma to become much more intense.

In order to analyze the distribution of the variables like density, temperature, velocity, data should be organized according to the occurrence of different results in each category. We have carried out a particle-in-cell (PIC) simulation which is depicted in figure (\ref{Fig:2}) and figure (\ref{Fig:2a}) and the procedure has been discussed in the appendix section. Plasma waves are formed by the free energy imparted to the system by the inward streaming solar wind electrons, and due to this particles starts disappearing from the simulation box. Thereafter, a particle starved region in phase-space emerge. The relatively moving particles pile over each other. Both components exist side by side to the right, form a distribution of particles that is unstable for the aforesaid instability under the condition that threshold velocity is much lower than relative velocity. The instability due to relative propagating particles in the inhomogeneous plasma becomes saturated, generating phase space starved pockets as particles are trapped by an increasingly unstable mode \cite{goldman1999nonlinear}. In the phase space simulation, the process takes place spatially and temporally as well. We investigated the efficiency of emerging two-stream instability of interacting solar wind ions with Venusian ionospheric plasma. The conservation of streaming energy for this case using the Poisson equation is obeyed. For time scale t=0 to 50 sec, the growing of two-stream instability has been checked for different initial conditions such as relative velocities and densities to study the regions where the energy conservation is obeyed. We have checked the energy conservation with relative velocity, 4.5 $km/s$ and density around 4 $cm^{-3}$. We have checked the ratio of energy of electrons to the initial energy with time scale (t). Initially the energy is zero. Then the instability sets in and the energy increases. Eventually the system comes to a dynamic equilibrium at a time scale, $t\approx30 s$, and the energy becomes constant at around 2.14 eV. 
When the accelerated stream of particles interacts with the stationary component of the electrons which were reflected further, a phase space trench forms toward the left of the box. It has been found that an electron-depleted region drifts away from the center of the phase space with increasing time.
\begin{table}[H]
\def\tablename{Data Table}
	\caption{Parameters for the Instability and the ion energization} 
	\centering 
	\begin{tabular}{p{4cm} c c} 
		\hline\hline 
		Parameter & Notation & Value \\ [0.5ex] 
		\hline 
		Density of SW & $n_{sw}$ & $5 ~cm^{-3}$  \\ 
		Velocity of SW & $v_{sw}$ & $350~km/s$\\
		Temperature of SW electrons & $T_{e}$ & $27~eV$ \\
		Thermal velocity of electrons & $v_{the}$ & $2500~km/s$ \\
		Planetary ion temperature & $T_{ion}$ & $0.3 - 0.4~eV$ \\
		Electron plasma frequency & $f_{e}$ & $9 ~ kHz$ \\
		Surface gravity of Venus & $g_{v}$ & $8.88~m/s^{2}$ \\
		\hline 
	\end{tabular}
	\label{table:nonlin} 
\end{table}

We have considered two situations here; one is when the velocity of threshold is lower than the relative velocity, and another is when the velocity of threshold is higher than the relative velocity. At first, all particles are able to flow smoothly. Figure (\ref{Fig:2}) shows the growth of the charges in phase space at different time. When the threshold velocity is much lower than relative velocity, all electrons become kinetic. Electrons subject to strong acceleration become kinetic. On other hand Figure (\ref{Fig:2a}) shows that the threshold velocity is large enough than the relative velocity then no instability arises. Now, we have already discussed that the threshold for instability, $\mathcal{F}(\omega_{s})>1$. We have used the data of the ionosphere of Venus. As we have seen two-stream instability is not seen everywhere in the ionosphere. there are some specific regions where the particles are able to get energy from such instability. In the next part of our study we have focused on that regions where ion-loss occurs. Our study lies in that velocity regime on which two-stream instability leads to a strong acceleration of planetary ions which are very energetic in the range of suprathermal energies \cite{sarkar2020formation}. Some newer simulation studies based on homotopy \cite{chandra2021evolution, ghosh2021resonant, ballav2021plasma, das2021electron, das2022semi} could provide some additional information.

\subsection{A Morphology of Ion Escape Process}
In different regions of the ionosphere, the average energies of the ions are different that corresponds to different fluxes of particle escape. 

\begin{figure}[H]
	\centering{\includegraphics[width=7.2cm, height=5.5cm]{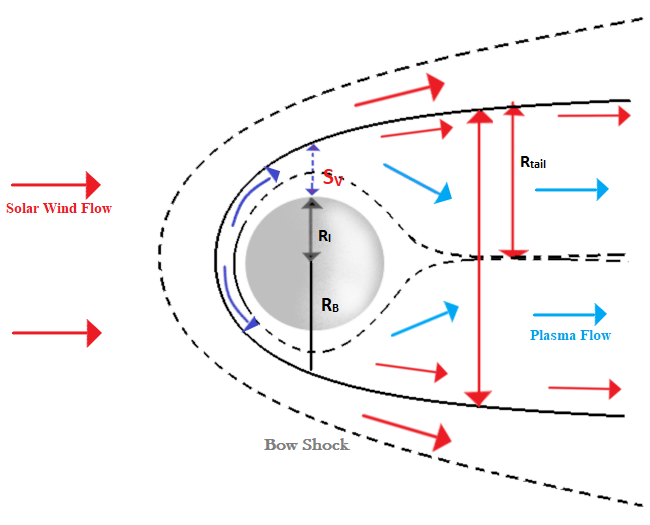}}
	\caption{A schematic diagram of energy-momentum transfer within deep magneto-tail region. Colored arrows indicate the flow of the SW, flow of planetary plasma.}
	\label{Fig:3}
\end{figure}

\noindent The maps of the measured fluxes are shown in the following Figure (\ref{Fig:4}) for the ions $H^+$, $O^+$ at the tail of the Venus plasma sheet \cite{lammer2006loss, rong2014morphology}. A schematic section is given in Figure (\ref{Fig:3}).

\begin{figure*}[ht]
	\centering{\includegraphics[height=5cm, width=17.5cm]{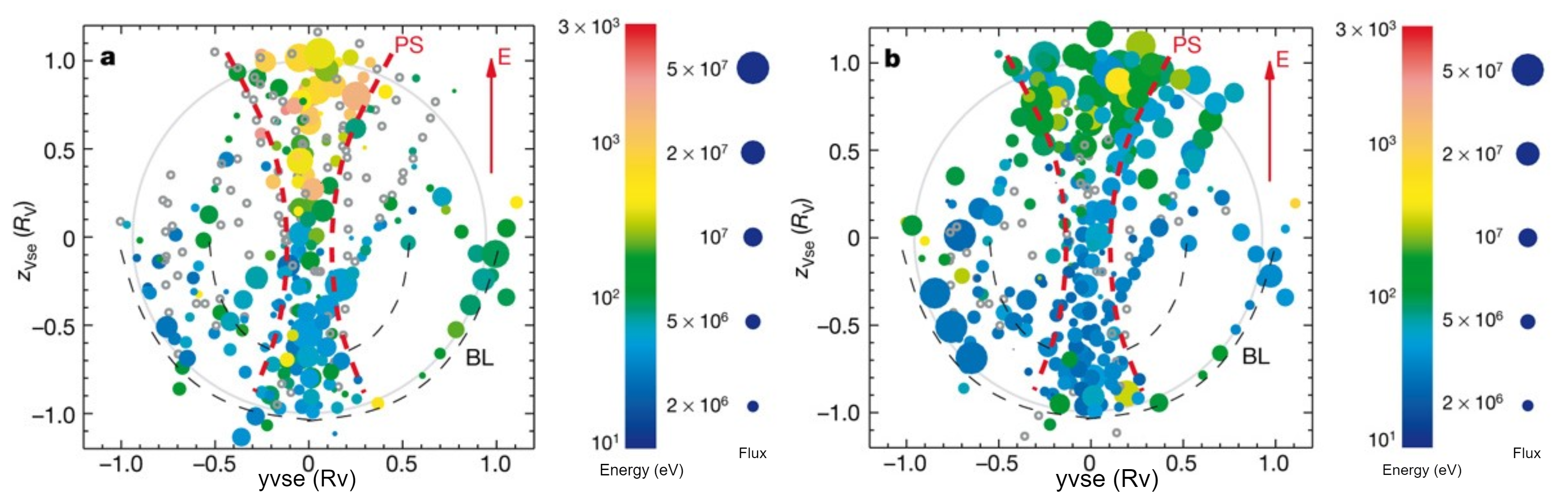}}
	\caption{Measured flux map of
		(a) $O^{+}$, (b) $H^{+}$ fluxes within the magneto tail area of Venus (obtained by Barabash et al.)\cite{barabash2007loss} in "Venus–solar–electrical coordinate"\cite{chai2015solar}. Colour map indicates the energy (eV) and dots are flux $\left(cm^{-2}s^{-1}\right)$}
		\label{Fig:4}
\end{figure*}

This section focuses on possible regions of ion escape from the observations of VEX, specifically from the instrument ASPERA-4. We represented the ion escape over the aforesaid velocity regime for two-stream instability. New patch retrieved from PDS-PPI IGPP UCLA \cite{brecht2021explanation,brecht2021data} extending upon ion densities and magnetic field have been used here. 

It has been seen that ions that escape slowly carry the higher mass loss. Like, if we consider for a while, that $O^+$ ions flow with velocity 'v' comprise that streaming instability, then it is the cause for supplying the velocity for escape. Thermal velocity is very near to the velocity of escape. The $\Phi_{max}$ (maximum flux) will be large than solar wind (SW) flux by the factor of that velocity v \cite{bingham2010modified}. We assumed that planetary ions are escaping across the plasma sheet, therefore the area where the exchange of momenta starts is given by,
\begin{equation}
	\centering
	S_V = \pi(R_B^2-R_I^2)
\end{equation} 
$R_B$ and $R_I$ are the radii of the boundaries where exchange of momenta starts and Venusian ionosphere respectively. Estimated flux, $\Phi_V\sim \frac{m_{SW}v_{SW}}{m_iv}\Phi_{SW}$
where, the flux of the solar wind is $\Phi_{SW}$ and $\Phi_v$ is the flux due to ions \cite{futaana2017solar}.

\subsection{A Possible Way of the Particle Escape Process}

A simple estimation of the flux escaping from Venus caused by the solar wind stands on momentum conservation. It has been found that ions like $O^{+}$ which are comparatively heavier than $H^{+}$ ions remain uninfluenced by the magnetic field. This causes an electric field to appear that extracts and also accelerate the planetary ions like $H^{+}, O^{+}$ to reach quasi-neutrality condition\cite{jarvinen2009oxygen}.
A cross-sectional view of flux measurement of ions like oxygen and hydrogen in the Venusian magnetotail region in "Venus-Solar-Electric coordinate system" \cite{chai2015solar} is shown in Figure (\ref{Fig:4}). The vertical red arrow defines the electric field.

\begin{figure}[H]
	\centering{\includegraphics[width=7.2cm, height=10.4cm]{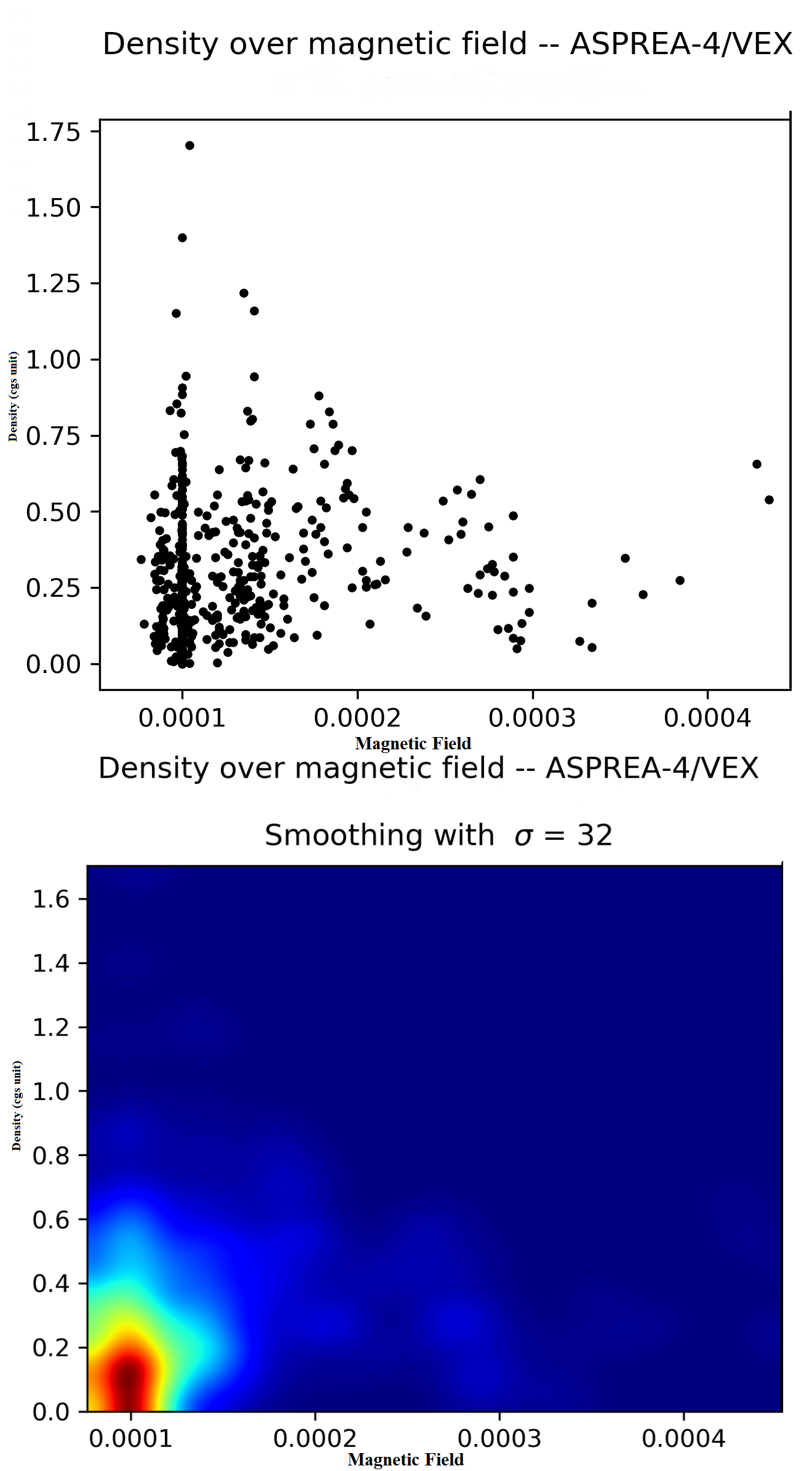}}
	\caption{Density of Oxygen like heavy ions over z-component of magnetic field plot and heatmap for Venus plasma ionosphere}
	\label{Fig:5}
\end{figure}

Density over magnetic field distribution was shown in Figure (\ref{Fig:5}). Higher density regions are slightly experienced by induced magnetic field. The thermal leakage of lighter ions, mostly $H^+$ ions, is quite dominant in Venus \cite{salem2020ionospheric}. On other hand, heavier atoms like that of Oxygen ($O^{+}$) follow the mechanism, $$O_{2}^{+}+e^{-}\rightarrow O^{*} (normal) + O^{*} (energetic)$$ and energy is supplied by two-stream instability for escape. Our investigation therefore can only explain ion-loss within the energy range 10 to 100 eV.

Hartle and Grabowsky \cite{hartle1990upward, hartle1993light} have shown that the lighter ion streams in the Venus ionospheric zone are caused by the magnetic and the electrical pressure gradients given by: 
\begin{equation}
	~q n_{e}\mathbf{E}=-\nabla P_{e} - \nabla \frac{B_{z}^{2}}{2\mu_{0}}
\end{equation}

\noindent We see that planetary protons or lighter ions $(H^{+})$ move quite faster than the comparably heavier ions $(O^{+})$ which are still much slower than ions in the solar wind flow. Bulk speeds of oxygen and hydrogen ions are grouped around, $V_{O^{+}}=V_{H^{+}}\times\sqrt{\frac{m_{H^{+}}}{m_{O^{+}}}}$ implying that the relation between the possible velocities of escape for particles with approximately same energy. In an effect, it may be concluded though subject to different electric fields, the ion's energy gain of similar order is obtained through the exchange of momentum \cite{ramstad2013phobos}. The relative velocity regime associated with our simulation strictly relates to the flux map in figure (\ref{Fig:4}). This shows a correspondence between simulation and particle escape scenarios.

The flux map showing the sky blue bubbles in figure (\ref{Fig:4}) ($10^1 - 10^2~eV$), are the regions where the instability is dominating. In this regions our PIC code is forming the instability. 
The particles can be out of the ionosphere of the planet if the ion velocity is greater than the escape velocity. Using the condition $\mathcal{F}(\omega_{s})>1$ we had got the threshold velocity to be :
\begin{equation}
    v_{th} \equiv v=\frac{\omega_{pe}}{k}\left[1+\left(\frac{m_{e}}{m_{p}}^{1/3}\right)\right]^{3/2}
\end{equation}
From this the Threshold Energy ($E_{th}$) comes out to be about 5.0661 $eV$ and using the fact that the extra energy is due to instability is about 2.138 $eV$. The escape energy of oxygen ions is nearly 8.7 eV. From the cross-sectional view, the energies of oxygen ions ranges from 10-100 eV, and the excess energy can be estimated using the threshold. We got extra energy of about 2 to 4 eV, which is enough to escape from the ionosphere.

So, the fundamental problem of depletion of the Hydrogen and Oxygen ions from the atmosphere can be directly related to the energization of ions through two-stream instability induced by the solar wind. The streaming instability energizes the ions, pushing them to velocities higher than the escape velocity, causing the loss of planetary ions.

Interestingly the rate of escape of the ions in the Venusian ionosphere, estimated by the various authors \cite{yadav2020plasma, luhmann2006venus,fedorov2011measurements} based on the observations by ASPERA-4 and in distinct energy spectra are quite similar in their form and is in tune with this present investigation. It may be also possible that some shock like nonlinear resonant mechanisms can accelerate the process \cite{sarkar2020resonant, chandra2021self, sarkar2022homotopy, maiti2021study, sahoo2021quantum}. In this respect some recent works can provide some insight \cite{mukhopadhyay2021electrostatic, goswami2021quantum}.

\section{Appendix}
We have designed a Particle-in-cell (PIC) simulation Figure (\ref{Fig:2}) to demonstrate the phase space mechanism and two-stream instability. We have framed a python code to simulate the instability. At the beginning we calculate the acceleration on each particle due to the field then compute the electron number density on the mesh by placing the particles into the 2 nearest bins (j and j+1, with proper weights) and then normalized them. Next we applied a periodic boundary condition on them following which we solved the Poisson Equation and applied derivative to get the electric field and interpolated the grid value onto the locations of particle. Afterward, we run our defined functions in the main PIC simulation function with appropriate parameter values. They are constructed by two matrices to computer Gradient and Laplacian for $1^{st}$ and $2^{nd}$ derivatives. Finally, simulating in main loop with drift (and applied some periodic boundary conditions) we obtain the two-stream instability in phase space. Such instabilities can be explained elaborately in the light of studies carried out by some authors \cite{dey2022rogue, shilpi2022study, sarkar2022heliospheric, chandra2022bifurcation, manna2022formation} using dynamical systems \cite{sahoo2015dust, sarkar2017study, singh2017electron, chatterjee2021study, ballav2021non}. In Figure (\ref{Fig:4}) Flux map has been shown in "Venus solar electrical coordinate" using python this has been motivated by Barabash  et  al.\cite{barabash2007loss} Density map Figure (\ref{Fig:5}) of Hydrogen and Oxygen is plotted first and then smoothen using the sigmas in python numerically. 

\section{ORCID iDs}
\noindent Suman Dey \includegraphics[width=0.035\linewidth]{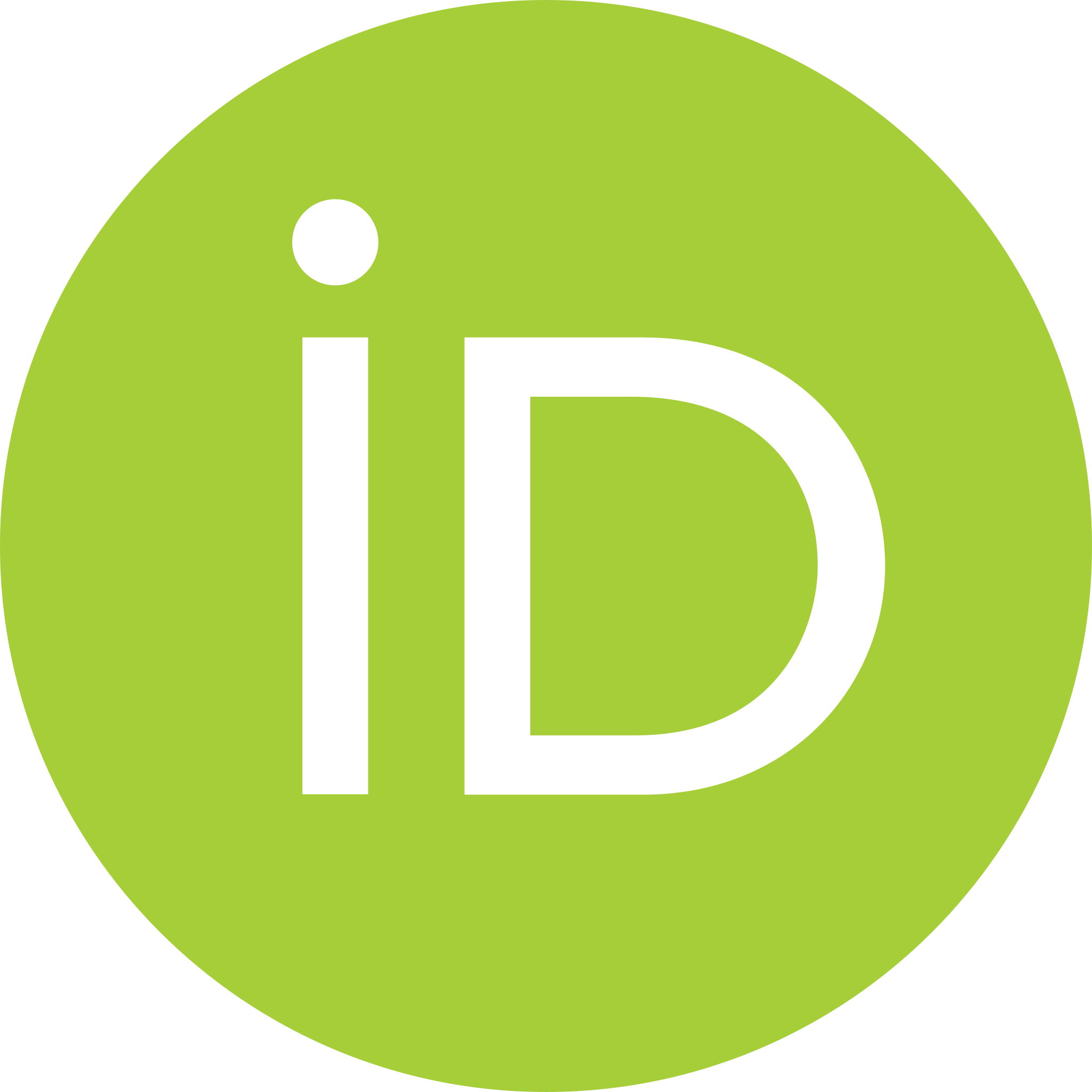} \href{https://orcid.org/0000-0003-1029-3807}{https://orcid.org/0000-0003-1029-3807}\\
Saptarshi Ghosh \includegraphics[width=0.035\linewidth]{ORCID.png} \href{https://orcid.org/0000-0001-8585-9743}{https://orcid.org/0000-0001-8585-9743}\\
Debjit Maity \includegraphics[width=0.035\linewidth]{ORCID.png} \href{https://orcid.org/0000-0001-8366-1872}{https://orcid.org/0000-0001-8366-1872}\\
Ayanabha De \includegraphics[width=0.035\linewidth]{ORCID.png} \href{https://orcid.org/
0000-0001-8845-184X}{https://orcid.org/0000-0001-8845-184X}\\
Swarniv Chandra \includegraphics[width=0.035\linewidth]{ORCID.png} \href{https://orcid.org/0000-0001-9410-1619}{https://orcid.org/0000-0001-9410-1619}\\

\section*{Acknowledgement}
 This work based on the data of the Venus Solar Wind Interactions, administered by Brecht, S. H.; Ledvina, S. A., with the identifier urn:nasa:pds:venus-solar-wind-interactions::1.0 which is archived at The NASA Planetary Data System (PDS), doi.org/10.17189/ \\1520598. This work makes use of data which are licensed under the terms of CC Attribution 4.0. These data were retrieved from PDS-PPI IGPP UCLA\cite{brecht2021explanation,brecht2021data} provided by European Space Agency (ESA). We also acknowledge the support received from the Institute of Natural Sciences and Applied Technology, Kolkata. The authors thank the reviewers whose comments and suggestions have helped to improve this manuscript. 



\end{document}